\documentclass[a4paper,11pt]{article}
\pdfoutput=1
\usepackage{jheppub}
\usepackage{color,bm,comment,braket}
\usepackage[T1]{fontenc}
\usepackage{comment}
\newcommand{\del}{\partial}
\newcommand{\beq}{\begin{eqnarray}}
\newcommand{\eeq}{\end{eqnarray}}

\newcommand{\non}{\nonumber\\}
\newcommand{\tr}{\mathop{\mathrm{tr}}}
\newcommand{\SU}{\text{SU}}
\newcommand{\U}{\text{U}}
\newcommand{\rmi}{\text{i}}
\newcommand{\rme}{\text{e}}
\newcommand{\rmd}{\text{d}}

\begin{document}

\title{
Domain-wall Skyrmion phase in a rapidly rotating QCD matter
}
\author[a,b]{Minoru Eto}
\emailAdd{meto@sci.kj.yamagata-u.ac.jp}
\affiliation[a]{%
Department of Physics, Yamagata University, 
Kojirakawa-machi 1-4-12, Yamagata,
Yamagata 990-8560, Japan}
\affiliation[b]{Research and Education Center for Natural Sciences, Keio University, 4-1-1 Hiyoshi, Yokohama, Kanagawa 223-8521, Japan}

\author[c,d,b]{Kentaro Nishimura}
\affiliation[c]{
International Institute for Sustainability with Knotted Chiral Meta Matter(SKCM$^2$), Hiroshima University, 1-3-2 Kagamiyama, Higashi-Hiroshima, Hiroshima 739-8511, Japan
}
\affiliation[d]{KEK Theory Center, Tsukuba 305-0801, Japan}
\emailAdd{nishiken@post.kek.jp}

\author[e,b,c]{and Muneto Nitta}
\emailAdd{nitta@phys-h.keio.ac.jp}
\affiliation[e]{Department of Physics, Keio University, 4-1-1 Hiyoshi, Kanagawa 223-8521, Japan}

\abstract{
Based on the chiral perturbation theory 
at the leading order, 
we show the presence of 
a new phase in rapidly rotating QCD matter with two flavors, that is 
a domain-wall Skyrmion phase.
Based on the chiral Lagrangian  with 
a Wess-Zumino-Witten (WZW) term responsible for 
the chiral anomaly and chiral vortical effect, 
it was shown that the ground state is 
a chiral soliton lattice(CSL) 
consisting of a stack of 
$\eta$-solitons 
in a high density region under rapid rotation. 
In a large parameter region, 
a single $\eta$-soliton decays 
into a pair of non-Abelian solitons, 
each of which carries 
${\rm SU}(2)_{\rm V}/{\rm U}(1) \simeq {\mathbb C}P^1 
\simeq S^2$ moduli 
as a consequence of the spontaneously broken 
vector symmetry ${\rm SU}(2)_{\rm V}$.
In such a non-Abelian CSL, 
we construct 
the effective world-volume theory of a single non-Abelian 
soliton to obtain
a $d=2+1$ dimensional ${\mathbb C}P^1$ model with a topological term 
originated from the WZW term. 
 We show
that when the chemical potential is larger than 
a critical value, 
 a topological lump supported by the second homotopy group $\pi_2(S^2) \simeq {\mathbb Z}$ has negative energy  
 and is  spontaneously created, 
 implying the domain-wall Skyrmion phase.
 This lump corresponds in the bulk to a Skyrmion
 supported by the third homotopy group $\pi_3[ {\rm SU}(2)] \simeq {\mathbb Z}$
 carrying a baryon number. 
 This composite state is called a domain-wall Skyrmion, 
 and 
  is stable even in the absence of the Skyrme term.
 An analytic formula for the effective nucleon mass 
 in this medium is obtained as
 $4\sqrt{2}\pi f_{\pi}f_\eta/m_{\pi} \sim 1.21$ GeV  with 
   the decay constants $f_{\pi}$ and $f_\eta$ of 
   the pions and $\eta$ meson, respectively, and 
   the pion mass $m_{\pi}$, 
which is surprisingly close to the  nucleon mass in the QCD vacuum.
}

\preprint{YGHP-23-04, KEK-TH-2570}

\maketitle

\section{Introduction}

Quantum Chromodynamics (QCD) is
the fundamental theory of the strong interaction 
described by quarks and gluons. 
QCD at extreme conditions such as 
high baryon density, strong magnetic field, 
and rapid rotation 
has been paid much attention 
since it is relevant for neutron star interiors 
and heavy-ion collisions 
\cite{Fukushima:2010bq}.
Lattice QCD cannot be extended to finite baryon density because of the notorious sign problem.
Instead, 
at least at low energy, 
the chiral Lagrangian or 
 the chiral perturbation theory (ChPT) 
 offers a powerful tool  
 since the theory is thoroughly determined by symmetry
up to some constants, the pion's decay constant, 
quark masses, and so on 
\cite{Scherer:2012xha,Bogner:2009bt}. 
When 
the chiral symmetry 
mixing different species of 
quarks (up-quarks, down-quarks and so on) 
is spontaneously broken, 
there appear 
 Nambu-Goldstone(NG) bosons or pions.
 Thus, the low-energy dynamics can be 
 described by the aforementioned ChPT. 

One of the most important extreme conditions for QCD 
is strong magnetic fields because of the interior of neutron stars 
and heavy-ion collisions.
In the presence of an external magnetic field,
the chiral Lagrangian is accompanied by 
the Wess-Zumino-Witten (WZW) term containing  
an anomalous coupling of the neutral pion $\pi_0$ to the magnetic field via the chiral anomaly 
\cite{Son:2004tq,Son:2007ny}
in terms of 
the Goldstone-Wilczek (GW) current \cite{Goldstone:1981kk,Witten:1983tw}.
It was determined to reproduce the so-called chiral separation effect (CSE) \cite{Vilenkin:1980fu,Son:2004tq,Metlitski:2005pr,Fukushima:2010bq,Landsteiner:2016led} in terms of the neutral pion $\pi_0$.
Then, 
at a finite baryon chemical potential $\mu_{\textrm{B}}$ under a sufficiently strong magnetic field, 
the ground state of QCD with two flavors (up and down quarks)  
was found to be 
a chiral soliton lattice (CSL) 
consisting of a stack of 
domain walls or solitons 
carrying a baryon number 
\cite{Son:2007ny,Eto:2012qd,Brauner:2016pko}. 
However, 
such a CSL state was found to be 
unstable against 
a charged pion condensation 
in a region of higher density and/or stronger  
magnetic field \cite{Brauner:2016pko}.
In such a region,
there appears a new phase,
the domain-wall Skyrmion phase
in which Skyrmions are created on top of the solitons 
in the ground state \cite{Eto:2023lyo,Eto:2023wul}.\footnote{
A possibility of an Abrikosov's vortex lattice was also proposed in the unstable region \cite{Evans:2022hwr}. 
See also a recent paper \cite{Evans:2023hms}.
} 
To show this,
the effective world-volume theory on a single soliton was constructed as an O(3) sigma model 
or the ${\mathbb C}P^1$ model with topological terms induced from the WZW term. Then, topological lumps (or baby Skyrmions) 
supported by $\pi_2({\mathbb C}P^1)\simeq {\mathbb Z}$
on the world volume, 
corresponding to 3D Skyrmions supported by 
$\pi_3[{\rm SU}(2)]\simeq {\mathbb Z}$
in the bulk point of view, 
appear in the ground state for a sufficiently large chemical potential. 
The composite states of a domain wall and Skyrmions are called 
domain-wall Skyrmions. 
Such domain-wall Skyrmions 
were previously proposed and studied in field theory 
\cite{Nitta:2012wi,Nitta:2012rq,
Gudnason:2014nba,Gudnason:2014hsa,Eto:2015uqa,Nitta:2022ahj}.\footnote{
The term ``domain-wall Skyrmions'' was first used 
in ref.~\cite{Eto:2005cc}
for 
Yang-Mills instantons 
absorbed into a domain wall, 
which can be described as Skyrmions in the domain-wall effective theory.
This term is different from ours.
} 
Domain-wall Skyrmions of a 2+1 dimensional version were also proposed in field 
theory 
\cite{Nitta:2012xq,Kobayashi:2013ju,Jennings:2013aea}
and  have been recently 
observed  experimentally 
\cite{Nagase:2020imn,Yang:2021} 
in chiral magnets \cite{PhysRevB.99.184412,KBRBSK,Ross:2022vsa,Amari:2023gqv,Amari:2023bmx} (see also Refs.~\cite{PhysRevB.102.094402,Kim:2017lsi,Lee:2022rxi}).

Another important extreme condition for QCD 
that we focus on in this paper 
is a rapid rotation. 
Quark-gluon plasmas produced in non-central heavy-ion collision experiments at the Relativistic Heavy Ion Collider (RHIC) reach the largest vorticity observed thus far, of the order of $10^{22}/{\rm s}$ \cite{STAR:2017ckg,STAR:2018gyt}. 
This has triggered significant attention to rotating QCD matter in recent years 
\cite{Chen:2015hfc,Ebihara:2016fwa,Jiang:2016wvv,Chernodub:2016kxh,Chernodub:2017ref,Liu:2017zhl,Zhang:2018ome,Wang:2018zrn,Chen:2019tcp,Chernodub:2020qah, Chernodub:2022veq, Huang:2017pqe,Nishimura:2020odq,Chen:2021aiq,
Eto:2021gyy}.
In particular, 
similar but different type of CSL appears in QCD 
at finite density under rapid rotation 
instead of a strong magnetic field.\footnote{
Now CSLs appear in various situations in QCD: 
CSLs  
under thermal fluctuation \cite{Brauner:2017uiu,Brauner:2017mui,Brauner:2021sci,Brauner:2023ort},  
quantum nucleation of CSLs 
\cite{Eto:2022lhu,Higaki:2022gnw}
and quasicrystals \cite{Qiu:2023guy}.
Possible relations between Skyrmion crystals 
at zero magnetic field 
and 
the CSL phase was discussed in 
Refs.~\cite{Kawaguchi:2018fpi,Chen:2021vou,Chen:2023jbq}. 
}
The anomalous term for the $\eta'$ meson  was obtained 
\cite{Huang:2017pqe,Nishimura:2020odq} 
by matching with 
the chiral vortical effect (CVE) \cite{Vilenkin:1979ui,Vilenkin:1980zv,Son:2009tf,Landsteiner:2011cp,Landsteiner:2012kd,Landsteiner:2016led} 
in terms of mesons.
Although 
the full WZW term is not known 
unlike the case of the magnetic field, 
this CVE term is sufficient to yield 
a CSL phase made of the $\eta'$ meson 
as the ground state 
in a certain parameter region 
\cite{Huang:2017pqe,Nishimura:2020odq,Chen:2021aiq},  
instead of that of the neutral pion $\pi_0$ 
in the case of the magnetic field.\footnote{
A different type of inhomogeneity of rotating matter
was also discussed in refs.~\cite{Chernodub:2020qah, Chernodub:2022veq}.
}
In the two-flavor case, there appears 
a CSL phase made of the $\eta$ meson. 
However, it was shown in ref.~\cite{Eto:2021gyy} that 
in a large parameter region,
a single $\eta$-soliton 
energetically decays into a pair of non-Abelian solitons, 
around which the neutral pion condensation occurs.
A single non-Abelian soliton 
spontaneously breaks the vector symmetry 
${\rm SU}(2)_{\rm V}$ into its U$(1)$ 
subgroup, resulting in NG modes
${\rm SU}(2)_{\rm V}/{\rm U}(1) \simeq {\mathbb C}P^1 
\simeq S^2$ localized in the vicinity of the soliton.
Thus, 
as the case of a $\pi_0$ soliton 
in the magnetic field, 
each non-Abelian soliton carries 
${\mathbb C}P^1$ moduli and are 
called non-Abelian sine-Gordon solitons
\cite{Nitta:2014rxa,Eto:2015uqa} 
(see also refs.~\cite{Nitta:2015mma,Nitta:2015mxa,Nitta:2022ahj}).
In a lattice of non-Abelian solitons, 
the ${\mathbb C}P^1$ modes at 
two neighboring solitons 
repel each other, and thus they are antialigned.
The lattice behaves as a Heisenberg anti-ferromagnet, 
in which we call one soliton an up-soliton 
and its neighbors down-solitons, 
and then up and down solitons appear alternately.
Such a non-Abelian CSL can be classified  
 into the two cases, the deconfined and dimer phases.
In the deconfined phase, 
an up-soliton and down-soliton repel each other, 
and they are separated with the equal distance.
In the dimmer phase, 
they attract each other at large distances 
and repel at short distances, 
and thus constitute a molecule. 
The lattice can be regarded as a lattice of molecules.
On the other hand, in the confined phase, 
the up and down-solitons 
attract and are completely overlapped to become 
$\eta$ solitons.
This is the previously known $\eta$-CSL 
\cite{Nishimura:2020odq}, 
in which 
the vector symmetry SU$(2)_{\rm V}$ is unbroken, 
and no soliton carries ${\mathbb C}P^1$ modes.

In this paper, we establish the presence of 
a new phase in rapidly rotating QCD matter, namely 
a domain-wall Skyrmion phase inside 
the non-Abelian CSL, 
similar to the case of a strong magnetic field. 
Within non-Abelian CSLs, either in the deconfined or dimer phase, the vector symmetry 
SU$(2)_{\rm V}$ is spontaneously broken 
into its U$(1)$ 
subgroup, thus being accompanied by 
NG modes
SU$(2)_{\rm V}/{\rm U}(1) \simeq {\mathbb C}P^1 
\simeq S^2$ as mentioned above.
For our purpose we concentrate on 
a single non-Abelian soliton in 
the deconfined phase.
We construct the effective theory on a single non-Abelian soliton and obtain  
a ${\mathbb C}P^1$ model with a topological term 
originated from the WZW term.
It admits topological lumps (baby Skyrmions) ensured by the second homotopy group $\pi_2(S^2) \simeq {\mathbb Z}$ 
\cite{Polyakov:1975yp}.
We find that when the chemical potential is larger than 
a critical value, 
the topological lumps have negative energy 
due to the WZW term.
 The lumps on the non-Abelian soliton are Skyrmions supported by the third homotopy group 
 $\pi_3(S^3) \simeq {\mathbb Z}$ 
 in the bulk point of view \cite{Eto:2015uqa}, and they carry 
 baryon numbers.
 This implies the domain-wall Skyrmion phase 
 in which lumps are spontaneously created 
 in the ground state,
 where the both chiral solitons and Skyrmions carry baryon numbers.
  The lump energy is obtained as
  $4\sqrt{2}\pi f_{\pi}f_\eta/m_{\pi}$
which can be interpreted as 
 the effective nucleon mass 
 in this medium (inside a soliton 
 at finite density under rapid rotation)  
 and is evaluated as $\sim 1.21$ GeV. 
 This value is close enough the  nucleon mass 938 MeV in the QCD vacuum.

This paper is organized as follows.
In sec.~\ref{sec:NACSL}, we review non-Abelian CSLs.
In sec.~\ref{sec:EFT}, we construct the effective worldvolume theory of a single non-Abelian soliton in the deconfined phase of non-Abelian CSL.
In sec.~\ref{sec:SWSk}, we constuct domain-wall Skyrmions and find the presence of the domain-wall Skyrmion phase.
Sec.~\ref{sec:summary} is devoted to a summary and discussion.


\section{Non-Abelian chiral soliton lattices under rotation}
\label{sec:NACSL}
We focus on the phase in which the $\U(2)_{\textrm{L}}\times \U(2)_{\textrm{R}}$ chiral symmetry is spontaneously broken down.
The low-energy dynamics can thus be described by an effective field theory of the pions -- ChPT.
A $2\times 2$ unitary matrix represents the pion fields
\begin{gather}
    U = \Sigma \rme^{\rmi \chi_0} \,, 
    \quad 
    \Sigma = \rme^{\rmi \tau_a\chi_a} \,,
\end{gather}
where $\tau_a$ ($a=1,2,3$) are the Pauli matrices with the normalization $\tr(\tau_a\tau_b)=2\delta_{ab}$.
This field $\Sigma$ transforms under 
the $\SU(2)_{\textrm{L}}\times \SU(2)_{\textrm{R}}$ chiral symmetry as
    $\Sigma \to L\Sigma R^{\dag}$,
where $L$ and $R$ are $2\times 2$ unitary matrices,
while $\chi_0$ transforms under the axial $\U(1)_{\textrm{A}}$ symmetry as
$\chi_0 \to \chi_0 + 2\theta_0$.

Then, the effective Lagrangian at the leading order is 
($\mu=0,\cdots,3$)
\begin{align}
    \mathcal{L}_{\textrm{ChPT}}
    &= \frac{f_{\pi}^2}{4}g^{\mu \nu}\tr(\del_{\mu}U \del_{\nu}U^{\dag})
    - \frac{f_{\eta}^2-f_{\pi}^2}{8}g^{\mu \nu}\tr(U^{\dag}\del_{\mu}U)
    \tr(U^{\dag}\del_{\nu}U) \notag \\
    &+ \frac{Bm}{2}\tr(U+U^{\dag}-2\bm{1}_2)
    + \frac{A}{2}(\det U + \det U^{\dag}-2) \label{rot_Chiral_Lagrangian} \,,
\end{align}
where $f_{\pi}$ and $f_{\eta}$ are the decay constants of the pions and the $\U(1)_{\textrm{A}}$ singlet ($\eta$) meson , respectively, $m$ is the quark mass, 
and $A$ and $B$ are parameters that cannot be determined by symmetry alone.
The first and second terms are the kinetic terms of the $\chi_a$ and $\chi_0$, respectively.
The third term is the mass term of the mesons, stemming from the explicit chiral symmetry breaking due to the finite quark masses.
Then, the pion mass $m_\pi$ is related 
to the quark mass 
by the Gell-Mann--Oakes--Renner relation $Bm= f_\pi^2 m_\pi^2/4$.
The fourth term represents the QCD anomaly: $\U(1)_{\textrm{A}} \to \mathbb{Z}_4$.
The parameter $A$ gives an additional mass term for $\chi_0$ meson given by $\delta m_{\chi_0}^2 = A/f_{\eta}^2$. 
Here, 
$g_{\mu \nu}$ is a spacetime metric representing the rotating coordinates,
and $g^{\mu \nu}$ is its inverse:
\begin{gather}
    g_{\mu \nu}=
    \left(
    \begin{array}{cccc}
      1-\Omega^2(x^2+y^2) & \Omega y & -\Omega x & 0 \\
      \Omega y & -1 & 0 & 0 \\
      -\Omega x & 0 & -1 & 0 \\
      0 & 0 & 0 & -1
    \end{array}
    \right) \,, \label{rot_metric} \\
    g^{\mu \nu}=
    \left(
    \begin{array}{cccc}
      1 & \Omega y & -\Omega x & 0 \\
    \Omega y & \Omega^2 y^2-1 & -\Omega^2xy & 0 \\
   -\Omega x & -\Omega^2xy & \Omega^2x^2-1 & 0 \\
      0 & 0 & 0 & -1
    \end{array}
    \right) \,, \label{inverse_of_rot_metric}
\end{gather}
where $\Omega$ stands for the rotation velocity.

The external $\U(1)_{\textrm{B}}$ gauge field $A^{\textrm{B}}_{\mu}$ can couple to  $\Sigma$ through the GW current \cite{Goldstone:1981kk,Witten:1983tw}
\begin{gather}
    j_{\textrm{GW}}^{\mu}
    = -\frac{\epsilon^{\mu \nu \alpha \beta}}{24\pi^2} \tr
        L_{\nu}L_{\alpha}L_{\beta}
        \,,
\label{eq:GW}
\end{gather}
where we have introduced the standard notation $L_{\mu}\equiv \Sigma \del_{\mu}\Sigma^{\dag}$ and $R_{\mu}\equiv \del_{\mu}\Sigma^{\dag}\Sigma$.
Then, the effective Lagrangian coupling to $A^{\textrm{B}}_{\mu}$ can be written as
\begin{gather}
    \mathcal{L}_{\textrm{GW}} = A^{\textrm{B}}_{\mu}j_{\textrm{GW}}^{\mu}
    \subseteq{\mathcal{L}_{\textrm{WZW}}}
    \label{effective_Lagrangian_GW} \,.
\end{gather}
In order to introduce finite baryon chemical potential $\mu_{\textrm{B}}$,
we choose the $\U(1)_{\textrm{B}}$ gauge field as $A^{\textrm{B}}_{\mu}=(\mu_{\textrm{B}}, \bm{0})$.
This is a WZW term but 
the full expression for rotation $\mathcal{L}_{\textrm{WZW}}$
is not known thus far, 
in contrast to the case of the magnetic field 
in which case the full expression is known 
\cite{Son:2004tq,Son:2007ny}.
In the external electromagnetic field,
the gauge-invariant and conserved baryon current can be derived by the ``trial and error'' $\U(1)_{\textrm{em}}$ gauging \cite{Witten:1983tw}.
The coupling of $\U(1)_{\textrm{em}}$ gauged GW current to $A^{\textrm{B}}_{\mu}$ is calculated as \cite{Goldstone:1981kk,Son:2007ny}
\begin{gather}
    \tilde{\mathcal{L}}_{\textrm{WZW}} = A^{\textrm{B}}_{\mu}\tilde{j}^{\mu}_{\textrm{GW}} \,, \\
    \tilde{j}^{\mu}_{\textrm{GW}}
    =-\frac{1}{24\pi^2}\epsilon^{\mu \nu \alpha \beta}\{
    \tr(L_{\nu}L_{\alpha}L_{\beta})
    -3\rmi e \del_{\nu}[A_{\alpha}\tr(QL_{\beta}+QR_{\beta})]
    \} \,,
\end{gather}
The second term with $\Sigma=\rme^{\rmi \chi_3 \tau_3}$ becomes $e\bm{B}\cdot \bm{\nabla}\chi_3/(4\pi^2)$.
In fact, the above equation can be derived by reproducing the chiral separation effect (CSE) \cite{Vilenkin:1980fu,Son:2004tq,Metlitski:2005pr,Fukushima:2010bq,Landsteiner:2016led} in terms of $\chi_3$ meson.
This procedure is justified by the fact that the chiral anomaly coefficient determines the transport coefficient of the CSE \cite{Son:2009tf,Neiman:2010zi}.
Therefore, the anomaly matching of the CSE can derive the part of $\tilde{\mathcal{L}}_{\textrm{WZW}}$.
Unfortunately, the GW current is already invariant under general coordinate transformations, so the method applied to the electromagnetic field cannot be used.
Hence, the method to derive the full expression is not known.
However, when applying this method to a rotating system,
it is evident that at least the following terms exist. 
QCD at finite baryon chemical potential under global rotation is known to exhibit the anomalous current in the direction of the rotation, which is the so-called CVE \cite{Vilenkin:1979ui,Vilenkin:1980fu,Landsteiner:2011cp,Landsteiner:2012kd,Landsteiner:2016led} :
\begin{gather}
    \bm{j}_5 = 
    \langle \bar{q}\bm{\gamma}\gamma_5q \rangle
    =\frac{\mu_{\textrm{B}}^2}{\pi^2N_{\textrm{c}}}\bm{\Omega} \,,
\end{gather}
where $q$ is a quark field.
We note that the chiral anomaly determines the transport coefficient of CVE.
Therefore, due to the exactness of the chiral anomaly coefficient,
it must be reproduced in terms of the $\chi_0$ meson in the ChPT.
The anomaly matching for the CVE gives us the anomalous coupling of the $\chi_0$ meson to the rotation \cite{Huang:2017pqe,Nishimura:2020odq} :
\begin{gather}
    \mathcal{L}_{\textrm{CVE}}
    = \frac{\mu_{\textrm{B}}^2}{2\pi^2N_{\textrm{c}}}
    \bm{\Omega}\cdot \bm{\nabla}\chi_0 \,.\label{eq:CVE}
\end{gather}
Hereafter, 
we interpret that 
$\mathcal{L}_{\textrm{CVE}}$ 
is a part of 
$\mathcal{L}_{\textrm{WZW}}$.

To derive an effective Lagrangian,
we adopt a modification to the conventional power counting scheme in ChPT \cite{Brauner:2021sci}:
\begin{gather}
    \del_{\mu} \,, m_{\pi} \sim \sqrt{Bm} \,, A_{\mu} \,,
    \Omega \,,
    \in \mathcal{O}(p^1), \\
    A^{\text{B}}_{\mu} \sim \mu_{\text{B}} \in \mathcal{O}(p^{-1}).
\end{gather}
In this power counting, eq.~(\ref{effective_Lagrangian_GW}) is of order $\mathcal{O}(p^2)$, consistent with eq.~(\ref{rot_Chiral_Lagrangian}).
The sole appearance of $\mu_{\text{B}}$ in the WZW term in eq.~(\ref{effective_Lagrangian_GW}) permits the assignment of a negative power counting to $\mu_{\text{B}}$.
The effective field theory up to $\mathcal{O}(p^2)$ encompasses the terms in eq.~(\ref{eq:GW});
however, eq.~(\ref{eq:GW}) has been overlooked in prior studies of the CSLs under rotation.
For discussions related to the magnetic field,
we refer to our previous work \cite{Eto:2023lyo,Eto:2023wul}.
In the QCD vacuum,
the effects of the QCD anomaly are generally not suppressed.
Hence, we note that it is not feasible to incorporate the QCD anomaly's effects into ChPT (Of course, in the large-$N_{\textrm{c}}$ expansion, the effects of the QCD anomaly are of the order of $1/N_{\textrm{c}}$, allowing them to be treated perturbatively \cite{Witten:1979vv,Witten:1980sp}).
We underscore that an $\mathcal{O}(p^4)$ term, such as a Skyrme term, is unnecessary for our findings.

Our effective theory when ignoring the charged pions $\chi_{1,2}$ 
and assuming one-dimensional dependence in the $x^3$ coordinate 
reduces to 
\begin{gather}
    \frac{\mathcal{H}}{C}
    = \frac{1-\epsilon}{2}(\chi_3^{\prime})^2 + \frac{1}{2}(\chi_0^{\prime})^2
    + \sin{\beta} (1-\cos{2\chi_0}) + \cos{\beta} (1-\cos{\chi_0}\cos{\chi_3}) - S \chi_0^{\prime} \label{eq:bulk_Hamiltonian} \,,
\end{gather}
where 
the prime denotes a differentiation 
with respect to $x^3$ and 
we have introduced the following quantities,
\begin{gather}
\textrm{sin}\beta \equiv \frac{A}{C} \,, \qquad \textrm{cos}\beta \equiv \frac{2Bm}{C} \,, \qquad C \equiv \sqrt{A^2 + (2Bm)^2} \,.
\end{gather}
and dimensionless variables as follows:
\begin{gather}
    \zeta \equiv \frac{\sqrt{C}z}{f_{\eta}} \,, \qquad
    \epsilon \equiv 1 - \left(\frac{f_{\pi}}{f_{\eta}} \right)^2 \,, \qquad
    S \equiv \frac{\Omega \mu_{\textrm{B}}^2}{2\pi^2 N_{\textrm{c}}\sqrt{C}} \label{dimensionless_variables} \,.
\end{gather}
The third and fourth terms
 are the potential terms of the $\chi_0$ and $\chi_3$:
\begin{gather}
    \frac{V_{\textrm{pot}}}{C} = \cos \beta (1-\cos \chi_0 \, \cos \chi_3)
    + \sin \beta (1-\cos 2\chi_0) \, .
\end{gather}

For later convenience,
we introduce new fields defined as
\begin{gather}
    \chi_{\pm} \equiv \chi_0 \pm \chi_3 \,.
\end{gather}
In terms of $\chi_+$ and $\chi_-$,
the Hamiltonian is reduced as
\begin{align}
    \frac{\mathcal{H}}{C}
    &= \frac{1}{2}\left[
    \frac{1}{2}\left(\frac{\rmd \chi_+}{\rmd \zeta} \right)^2
    + (1-\cos{\chi_+}) - S\frac{\rmd \chi_+}{\rmd \zeta}
    \right]
    + \frac{1}{2}\left[
    \frac{1}{2}\left(\frac{\rmd \chi_-}{\rmd \zeta} \right)^2
    + (1-\cos{\chi_-}) - S\frac{\rmd \chi_-}{\rmd \zeta}
    \right] \notag \\
    & -\frac{\epsilon}{8}(\chi_+^{\prime})^2 - \frac{\epsilon}{8}(\chi_-^{\prime})^2
    + \frac{\epsilon}{4}\chi_+^{\prime}\chi_-^{\prime}
    - \sin \beta (1-\cos (\chi_+ + \chi_-))
    -S\chi_0^{\prime} \,.
\end{align}
Eq.~(\ref{eq:bulk_Hamiltonian}) gives the equation of motions as follows:
\begin{gather}
    \chi_+^{\prime \prime}
    - \frac{2-\epsilon}{2(1-\epsilon)}\cos \beta \, \sin \chi_+
    + \frac{\epsilon}{2(1-\epsilon)}\cos \beta \, \sin \chi_-
    -2\sin \beta \, \sin(\chi_++\chi_-) = 0 \,, 
    \label{eq:eom_chi+}\\
    \chi_-^{\prime \prime}
    - \frac{2-\epsilon}{2(1-\epsilon)}\cos \beta \, \sin \chi_-
    + \frac{\epsilon}{2(1-\epsilon)}\cos \beta \, \sin \chi_+
    -2\sin \beta \, \sin(\chi_++\chi_-) = 0 \,.
    \label{eq:eom_chi-}
\end{gather}

\begin{figure}[tb]
    \centering
    \includegraphics[width=13.0cm]{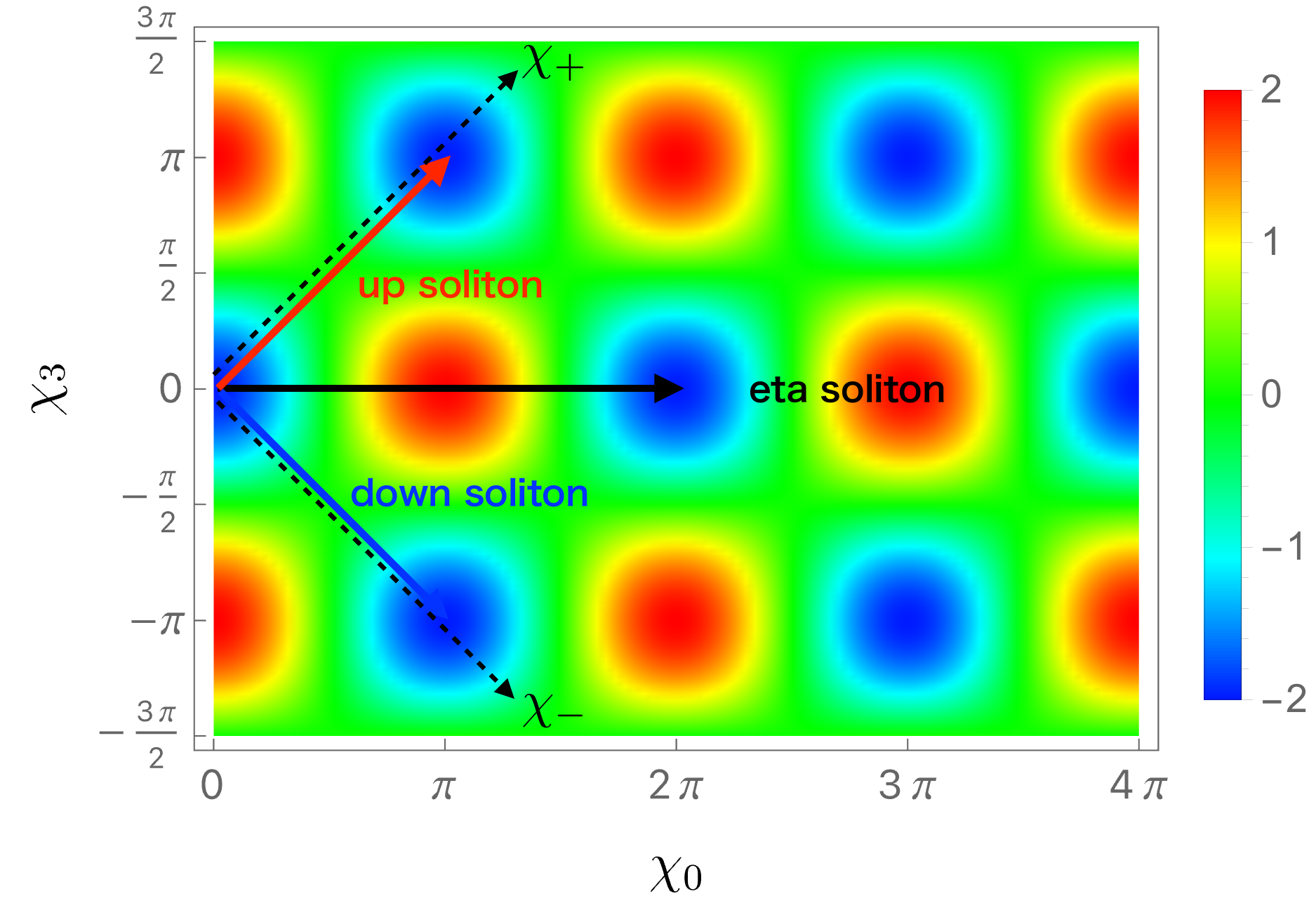}
    \caption{
   Single $\eta$ soliton and up and down non-Abelian solitons in the field space.
    The color denotes the height of the potential and 
    blues are vacua. 
    Fixing the state $(\phi_0,\phi_3)=(0,0)$ at $x^3=-\infty$, 
    the $\eta$ soliton denoted by the black arrow connects it to 
    $(\phi_0,\phi_3)=(2\pi,0)$ 
     the up soliton denoted by the red arrow connects it to 
    $(\phi_0,\phi_3)=(\pi,\pi)$ and 
     the down soliton denoted by the blue arrow connects it to 
    $(\phi_0,\phi_3)=(\pi,-\pi)$ at $x^3=+\infty$.
    }
    \label{fig:potential_and_kink}
\end{figure}

Let us first consider the case of $\epsilon=0$ and $\beta=0$.
The potential term at $\beta=0$ is sketched as fig.~\ref{fig:potential_and_kink}.
The configuration connecting $(0,0)$ and $(2\pi,0)$ is well-known as a single sine-Gordon soliton,
\begin{gather}
    \chi_0 = 4\tan^{-1}\rme^{\zeta-\zeta_0} \,,
    \label{eq:Abelian_CSL}
\end{gather}
which has the transnational moduli, $\zeta_0$.
On the other hand,
the configuration connecting $(0,0)$ and $(\pi,\pi)$ is given by
\begin{gather}
    U_+ = \textrm{diag}(u^{\rmi \theta},1) \,, \\
    \theta = 4\tan^{-1}\rme^{\zeta-\zeta_0} \,.
\end{gather}
which also has the translational moduli.
We note that this soliton  spontaneously breaks the $\SU(2)_{\textrm{V}}$ symmetry to a $\U(1)$ subgroup:
\begin{gather}
    U_+ \to gU_+g^{\dag} = U_+ \,, \\
    g=\rme^{\rmi \alpha \tau_3} \,.
\end{gather}
Therefore,
this soliton has not only translational moduli $\mathbb{R}$ but also $\SU(2)/\U(1) \cong {\mathbb C}P^1 \cong S^2$ moduli.
We call this soliton 
an up soliton.
The configuration connecting $(0,0)$ and $(\pi,-\pi)$ is given by the $\SU(2)_{\textrm{V}}$ transformation:
\begin{gather}
    U_- = \rme^{\rmi \pi \tau_1/2}U_+\rme^{-\rmi \pi \tau_1/2}
    = \textrm{diag}(1, u^{\rmi \theta}) \,,
\end{gather}
which is referred to as a down soliton.
The up soliton and down soliton 
are connected by 
the ${\mathbb C}P^1$ moduli.

\begin{figure}[tb]
    \centering
    \includegraphics[width=15.0cm]{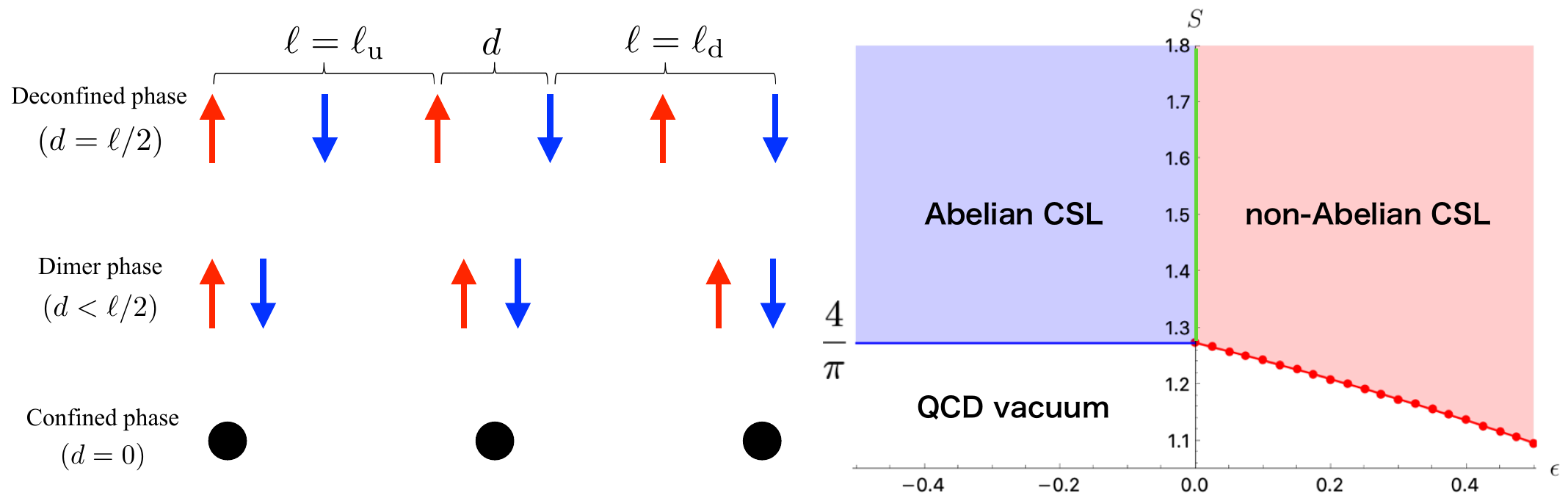}
    \caption{
    The schemetic picture of the three phases (left) and the phase diagram for $\beta =0$ (right). 
    (Left panel) The deconfined, dimer and confined phases in which up and down solitons repel, 
    form a molecule, and are completely overlapped, respectively.
    (Right panel) In this parameter choice, the dimer phase does not appear, and the whole non-Abelian CSL implies the deconfined phase. The gree line correspond to the noninteractive case 
    in which the up and down solitons do not interact each other and form lattices independently.
    }
    \label{fig:phase_diagram}
\end{figure}

Next, let us 
consider the effects of non-zero $\epsilon$ and $\beta$.
The term depending on $\epsilon$ is 
\begin{gather}
    -\frac{\epsilon}{2}(\chi_3^{\prime})^2
    = -\frac{\epsilon}{8}(\chi_+^{\prime})^2 - \frac{\epsilon}{8}(\chi_-^{\prime})^2
    + \frac{\epsilon}{4}\chi_+^{\prime}\chi_-^{\prime} \label{epsilon_potential} \,.
\end{gather}
Since $\chi_{+} (\chi_{-})$ has a peak at the center of the up (down) soliton,
the energy density of eq.~(\ref{epsilon_potential}) becomes lower when the up soliton and down soliton are separated.
Therefore, there is a repulsive (attractive) force between the up and down solitons due to the finite positive (negative) $\epsilon$.
As $\beta$ increases from $0$,
the $\chi_3$ dependence of the potential decreases.
Therefore, the up and down soliton overlap to become the ordinary sine-Gordon soliton.
The effect of finite $\beta$ is the attractive interaction between the up and down solitons.

From the preceding discussion, we identify three distinct cases concerning the arrangement of the up and down solitons: the deconfined phase, the dimer phase, and the confined phase.

\begin{enumerate}
    \item \textbf{Deconfined Phase}: If the repulsive force significantly exceeds the attractive force, the up and down solitons fully separate. This state is termed the \emph{deconfined phase}. For this condition, the relationship between the distance \(d\) between the up and down solitons and the distance \(\ell\) between the same type of soliton is given by \(d = \frac{\ell}{2}\).
    
    \item \textbf{Confined Phase}: Conversely, when the attractive force strongly prevails over the repulsive force, the up and down solitons overlap entirely, denoted as \(d = 0\). This case is referred to as the \emph{confined phase}.
    
    \item \textbf{Dimer Phase}: When the attractive and repulsive forces counteract each other equally, a molecular state of the up and down solitons forms, satisfying the condition \(0< d < \frac{\ell}{2}\).
\end{enumerate}

When $\epsilon < 0$ the inter-soliton force is attractive, so that the CSL is in the confined phase. Namely, the CSL is of the Abelian type with $\chi_3=0$ and $\chi_0$ is the same as that given in eq.~(\ref{eq:Abelian_CSL}).
Then the Hamiltonian reads
\begin{gather}
    \frac{\mathcal{H}}{C}
    = \frac{1}{2}(\chi_0^{\prime})^2
    + (1-\cos{\chi_0}) - S \chi_0^{\prime}
    = 4\, {\rm sech}^2 \zeta - 2S\,{\rm sech}\,\zeta\,.
\end{gather}
The tension (the mass per unit area) is given by integrating ${\cal H}$ over $z$ as
\begin{eqnarray}
    \sigma\big|_{\text{1-soliton}} = \frac{f_\eta}{\sqrt C} \int^\infty_{-\infty}  d\zeta\, {\cal H}
    = f_\eta\sqrt{C}\left(8 - 2\pi S\right)\,.
\end{eqnarray}
This becomes zero at $S = 4/\pi$, and therefore the Abelian CSL becomes the ground state for $S\ge4/\pi$.
This is shown in fig.~\ref{fig:phase_diagram}.
When $\epsilon > 0$, the inter-soliton force is repulsive and therefore the ground state is the non-Abelian CSL. There are no analytic solutions to eqs.~(\ref{eq:eom_chi+}) and (\ref{eq:eom_chi-}), so that we numerically solve them and complete the phase diagram for $\beta = 0$ as shown in fig.~\ref{fig:phase_diagram}. We emphasize that the critical velocity decreases for $\epsilon > 0$.

\section{Non-Abelian sine-Gordon soliton and its effective world-volume theory} \label{sec:EFT}
In this section,
we construct the effective field theory of the $\U(2)$ non-Abelian sine-Gordon soliton under rotation 
by using the 
moduli (Manton) approximation 
\cite{Manton:1981mp,Eto:2006pg,Eto:2006uw}.
On the phase transition line between the vacuum and the deconfined phases,
the single soliton enters the system alternately,
and their distance is infinite.
Then, we focus only on the up soliton:
\begin{gather}
    U_0 = \textrm{diag}(\rme^{\rmi \chi_+}, 1)
    = \rme^{\rmi(\chi_+/2 + \chi_+\tau_3/2)} \,.
\end{gather}
Of course, one can choose the down soliton,
but since they are infinitely apart, it is sufficient to choose one.
Considering a sufficiently small $\epsilon$,
we can approximate $\chi_+ \simeq 4\tan^{-1}e^{\zeta}$.

So far, we have neglected the charged pions.
Its general solution containing the charged pions can be obtained from $U_0$ by an $\SU(2)_{\textrm{V}}$ transformation,
\begin{gather}
    U = gU_0g^{\dag} \label{eq:general_solution} \,,
\end{gather}
where $g$ is an $\SU(2)$ matrix.
Since $g$ in eq.~(\ref{eq:general_solution}) is redundant with respect to a $\U(1)$ subgroup generated by $\tau_3$,
it takes a value in a coset space, $\SU(2)_{\textrm{V}}/\U(1)\simeq \mathbb{C}P^1 \simeq S^2$.
Together with the translational modulus $Z$, 
the single  sine-Gordon soliton has the moduli 
\begin{gather}
    \mathcal{M} \simeq \mathbb{R} \times \mathbb{C}P^1 \,.
\end{gather}
Such a soliton with non-Abelian moduli is called a non-Abelian sine-Gordon soliton \cite{Nitta:2014rxa,Eto:2015uqa}.

Let us parameterize the $\mathbb{C}P^1$ moduli by the homogeneous coordinates $\phi \in \mathbb{C}^2$ of $\mathbb{C}P^1$, satisfying \cite{Eto:2015uqa}
\begin{gather}
    \phi^{\dag}\phi = 1 \,, \quad\quad
    g \tau_3 g^{\dag}
    = 2 \phi \phi^{\dag} - \bm{1}_2 \,.
\end{gather}
In terms of $\phi$,
eq.~(\ref{eq:general_solution}) is represented as
\begin{gather}
    U = \exp(\rmi \chi_+ \phi \phi^{\dag})
    = \bm{1}_2 + (u-1)\phi \phi^{\dag} \,, 
    \label{eq:general-U}
\end{gather}
where we define $u\equiv \rme^{\rmi \chi_+}$.
Since $\mathbb{C}P^1 \simeq S^2$,
the moduli space is also parameterized by the three-component real vector $\bm{n}$ with the unit length condition, $|\bm{n}|=1$.
$\phi$ and $n_a$ are related by the following formula:
\begin{gather}
    n_a = \phi^{\dag}\tau_a\phi \,.
\end{gather}
The condition $\phi^{\dag}\phi=1$ is solved by using the inhomogeneous coordinate $f\in \mathbb{C}$ as follow:
\begin{gather}
    \phi = \frac{1}{\sqrt{1+|f|^2}}\left(
    \begin{array}{c}
    1\\
    f
    \end{array}
    \right) \,.
\end{gather}
Then, the up soliton corresponds to $n_3=1$ ($f=0$) and 
the down solitons to $n_3=-1$ ($f=\infty$).

Now,
we are prepared to formulate the low-energy effective theory for a single soliton using the moduli approximation \cite{Manton:1981mp,Eto:2006pg,Eto:2006uw}. Consider a single sine-Gordon soliton perpendicular to the $x^3$-coordinate.
In what follows, the moduli parameter $\phi$ will be treated as fields on the soliton's $2+1$-dimensional worldvolume.
However, we will not do the same for the translational modulus $Z$ as its transverse motion is not pertinent to our investigation.
By substituting 
eq.~(\ref{eq:general-U})
into ${\cal L}$,
we get
\begin{gather}
    \mathcal{L}_{\textrm{kin}}
    = \frac{f_{\pi}^2+f_{\eta}^2}{8}g^{\mu \nu}
    \del_{\mu}\chi_+\del_{\nu}\chi_+
    + \frac{f_{\pi}^2}{2}|1-u|^2g^{\mu \nu}(
    \phi^{\dag}\del_{\mu}\phi\phi^{\dag}\del_{\nu}\phi
    + \del_{\mu}\phi^{\dag}\del_{\nu}\phi
    ) \,, \label{eq:EFT1} \\
    \mathcal{L}_{\textrm{mass}}
    = (Bm+A)(-1 + \cos \chi_+) \,, \label{eq:EFT2}\\
    \mathcal{L}_{\textrm{GW}}
    = \frac{\mu_{\textrm{B}} q}{4\pi}|1-u|^2 \del_3\chi_+\,.\label{eq:EFT3}
\end{gather}
Here, $q$ is the baby Skyrmion (lump) charge density defined by
\begin{gather}
    q \equiv
     -\frac{\rmi}{2\pi}\epsilon^{ij}\del_i\phi^{\dag}\del_j\phi
     = \frac{1}{8\pi} \epsilon^{ij}{\bm n}\cdot (\partial_i {\bm n}\times\partial_j{\bm n}) 
= \frac{i}{2\pi} \tr\left(\left[\partial_{\bar z} {\cal P},\partial_z {\cal P}\right]{\cal P}\right),
\quad {\cal P} \equiv \phi\phi^\dagger
    \label{eq:pi2}
\end{gather}
The integration of which over the two-dimensional space 
defines the topological lump number
\begin{eqnarray}
    k = \int \rmd^2x\, q \in 
   \pi_2(S^2) \simeq \mathbb{Z}\,.
   \label{eq:lump-number}
\end{eqnarray}
In the last expression in eq.~(\ref{eq:pi2}), 
 ${\cal P}$ is a projection operator ${\cal P}^2 = {\cal P}$ \cite{Din:1980jg,Din:1980wh}.
The integrations of the above Lagrangian
(\ref{eq:EFT1}), (\ref{eq:EFT2}) and (\ref{eq:EFT3})
over the codimension $x^3$ give us the total effective world-volume theory of the non-Abelian sine-Gordon soliton:
\begin{align}
    \mathcal{L}_{\textrm{DW}} 
    & = \int \rmd x^3 ( \mathcal{L}_{\textrm{kin}} 
    + \mathcal{L}_{\textrm{mass}}
    +  \mathcal{L}_{\textrm{GW}}
    + \mathcal{L}_{\textrm{CVE}}
    ) \notag\\
    &= -\frac{8\sqrt{C}(f_{\pi}^2+f_{\eta}^2)}{8f_{\eta}}
    - \frac{4f_{\eta}(Bm+A)}{\sqrt{C}}
    + \frac{\mu_{\textrm{B}}^2\Omega}{2\pi N_{\textrm{c}}} \notag \\
    & + \frac{4f_{\pi}^2f_{\eta}}{\sqrt{C}}g^{\alpha \beta}(
    \phi^{\dag}\del_{\alpha}\phi \phi^{\dag}\del_{\beta}\phi
    + \del_{\alpha}\phi^{\dag}\del_{\beta}\phi
    ) \notag \\
    & +\mu_{\textrm{B}}q \,, \label{eq:EFT}
\end{align}
where $\alpha,\beta = 0,1,2$ and we have used the integration formulas
\begin{gather}
    \int_{-\infty}^{\infty}\rmd x\, \left[
    \frac{f_{\eta}^2+f_{\pi}^2}{8}(\del_3\chi_+)^2
    + (Bm+A)(1-\cos \chi_+)
    \right]
    = \frac{8\sqrt{C}(f_{\pi}^2+f_{\eta}^2)}{8f_{\eta}}
    + \frac{4f_{\eta}(Bm+A)}{\sqrt{C}} \,, \\
    \int_{-\infty}^{\infty}\rmd x\, |1-u|^2 = \frac{8f_{\eta}}{\sqrt{C}} \,, \\
    \int_{-\infty}^{\infty}\rmd x\, \frac{1}{4\pi}
    |1-u|^2\del_3 \chi_+ =1 \label{eq:Hamiltonian_of_DW} \,.
\end{gather}
The equations in the second line represent the tension of the domain wall.
The third and fourth lines in eq.~(\ref{eq:EFT}) 
 denote the ${\mathbb C}P^1$ theory in the rotating coordinates and lump charge, respectively.
 The constant terms in eq.~(\ref{eq:EFT}) are 
irrelevant in our study; thus we ignore them hereafter.
The Lagrangian can be rewritten in terms of the inhomogeneous coordinate $f$ as
\begin{align}
    \mathcal{L}_{\textrm{DW}}
    &= \frac{f_{\pi}^2f_{\eta}}{\sqrt{C}}g^{\alpha \beta}
    \frac{\del_{\alpha}f\del_{\beta}f^*}{(1+|f|^2)^2}
    +\mu_{\textrm{B}}q \,.
\end{align}

In order to determine the ground state, let us calculate the momentum conjugate for $\del_0f$ and $\del_0f^*$:
\begin{eqnarray}
    \pi_f &\equiv& \frac{\del \mathcal{L}_{\textrm{DW}}}{\del(\del_0f)}
    = \del_0f^* + \Omega(y\del_1f^*-x\del_2f^*) \,, \\
    \pi_{f^*} &\equiv& \frac{\del \mathcal{L}_{\textrm{DW}}}{\del(\del_0f^*)}
    = \del_0f + \Omega(y\del_1f-x\del_2f) \,.
\end{eqnarray}
Then, the Hamiltonian can be calculated as follows: 
\begin{align}
    \mathcal{H}_{\textrm{DW}}
    &= \pi_f\del_0f + \pi_{f^*}\del_0f^* - \mathcal{L}_{\textrm{DW}} \notag \\
    &= \frac{f_{\pi}^2f_{\eta}}{\sqrt{C}}\frac{
    \del_0f\del_0f^* + \del_i f\del_if^*
    }{(1+|f|^2)^2} 
    - \mu_{\textrm{B}}q
    \notag\\
    &- \frac{f_{\pi}^2f_{\eta} \Omega^2}{\sqrt{C}}
    \frac{
    (y^2\del_1f\del_1f^* + x^2\del_2f\del_2f^*)
    + xy(\del_1f\del_2f^* + \del_2f\del_1f^*)
    }{(1+|f|^2)^2} \,.
    \label{eq:EFT0}
\end{align}
The last term proportional to $\Omega^2$ is
at higher order ${\cal O}(p^4)$ 
which we will omit in the following.

\section{Domain-wall Skyrmion phase}
\label{sec:SWSk}

In this section, we 
construct topological lumps 
in the domain-wall world-volume theory, 
and show that they correspond to 
Skyrmions in the bulk, 
implying the domain-wall Skyrmion phase.
For this purpose, let us introduce the complex coordinate:
\begin{gather}
    w \equiv x + \rmi y \,, \qquad
    \bar{w} \equiv x -\rmi y \,.
\end{gather}
Using these coordinates,
we have  
\begin{align}
    \mathcal{H}_{\textrm{DW}}
    &= \frac{f_{\pi}^2f_{\eta}}{\sqrt{C}}
    \frac{|\del_wf|^2+ |\del_{\bar{w}}f|^2}{(1+|f|^2)^2} 
    - \mu_{\textrm{B}}q \label{eq:DW_Hamiltonian} \,.
\end{align}
When the Bogomol'nyi-Prasad-Semmerfield (BPS) equation for $k>0$  \cite{Polyakov:1975yp}
\begin{align}
 \del_{\bar{w}}f = 0    
\end{align}
or the anti-BPS equation for $k<0$ 
\begin{align}
    \del_wf=0
\end{align}
holds,  the energy (\ref{eq:DW_Hamiltonian}) saturates 
the minimum (the Bogomol'nyi bound) 
of the following inequality:
\begin{align}
   E_{\textrm{DW}} 
   = \int \rmd^2 x\, \mathcal{H}_{\textrm{DW}}
    &\geq \frac{4\pi f_{\pi}^2f_{\eta}}{\sqrt{C}}
    \left|\int \rmd^2 x\, q\right| 
    -  \mu_{\textrm{B}} \int \rmd^2 x\, q \label{eq:DW_Hamiltonian-min} \,.
\end{align}
Due to the second term, anti-BPS lumps have more energy 
than BPS lumps.
Let us consider BPS $k$-lump solution 
\cite{Polyakov:1975yp}
\begin{eqnarray}
        f = \frac{b_{k-1}w^{k-1}+\cdots+b_0}{w^k + a_{k-1}w^{k-1}+\cdots+a_0}\,,
        \label{eq:k-lump_solution}
\end{eqnarray}
with moduli $a_i,b_i \in {\mathbb C}$ ($i=0,\cdots k-1$).

Now, let us show the relation between 
the topological lumps 
in the domain-wall world-volume 
theory and Skyrmions in the bulk. 
To this end,
the baryon (Skyrmion) number 
$B$ in the bulk 
 taking a value in $\pi_3 [{\rm SU}(2)]\simeq {\mathbb Z}$ can be calculated 
in terms of the Maurer-Cartan one-form 
$R_i \equiv U^\dagger \del_i U$ 
as \cite{Eto:2015uqa}
\beq
B &=& \frac{1}{24 \pi^2} \int d^3x\  \epsilon_{ijk} \tr (R_iR_jR_k) \non
&=& -\frac{1}{8 \pi^2} \int  d^3x\  \tr \left[\left(\partial_1U^\dagger\partial_2U - \partial_2U^\dagger\partial_1U\right)  U^\dagger \partial_3 U \right] \non
&=& -\frac{1}{8\pi^2}\int dx^1dx^2\ \tr\left(
\left[\partial_1 {\cal P},\partial_2{\cal P}\right] {\cal P}
\right) \int dx^3\ |u-1|^2 u^*\partial_3u\non
&=&  \frac{i}{2\pi}\int dw d\bar w\ \tr\left(
\left[\partial_{\bar z} {\cal P},\partial_z{\cal P}\right] {\cal P}
\right) \times \frac{1}{2\pi} \int dx^3\  (1-\cos\theta) \partial_3\theta\non
&=& k ,
\eeq
with the lump number $k$ defined in eq.~(\ref{eq:lump-number}).
Therefore, we have found that  
$k$ topological lumps
on the non-Abelian sine-Gordon soliton 
carry a baryon number $k$ and
represent $k$ Skyrmions in the bulk.
This one-to-one correspondence between 
lumps on the soliton and 
Skyrmions in the bulk has a sharp  
contrast to the domain-wall Skyrmion in the strong magnetic field,  
in which case 
one lump in the domain wall 
corresponds to
two Skyrmions in the bulk.

\begin{figure}[t]
\begin{center}
\includegraphics[width=15cm]{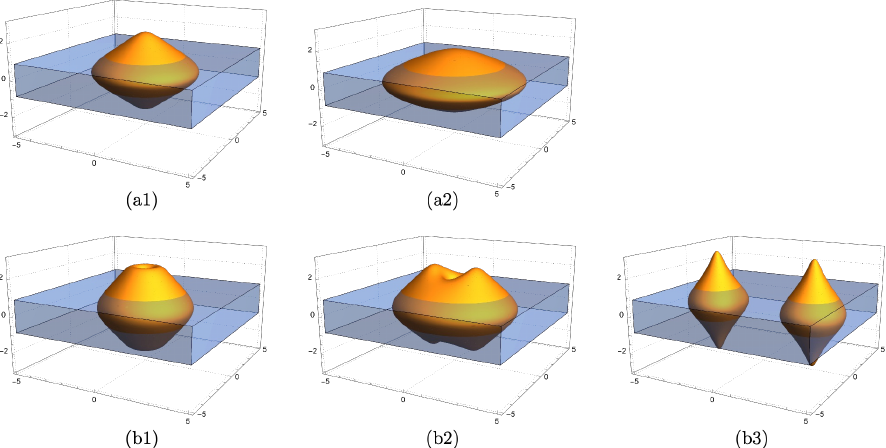}
\caption{
Three-dimensional configuration of $k=1$ and $k=2$ domain-wall Skyrmions for QCD at finite density under rapid rotation. An isosurface of the baryon density $(1/4\pi^2)\tr\left[R_1R_2R_3\right]$ is plot by orange surface and the blue region denotes a non-Abelian soliton. We take $m=1$. The top row shows $k=1$ with $|b_0|=1$ for (a1) and $|b_0|=2$ for (a2).
The bottom row corresponds to $f = b_0/(w-d)(w+d)$.
$(b_0,d)$ is taken as $(1,0)$ for (b1),
$(2,1)$ for (b2), and $(3,2)$ for (b3).
}
\label{fig:Go_stone}
\end{center}
\end{figure}

Fig.~\ref{fig:Go_stone} shows a three-dimensional configuration of a domain-wall Skyrmion. 
This can be compared with the case of that in strong magnetic field, 
in which a single lump has two peaks corresponding to two Skyrmions in the bulk.

Finally, let us evaluate the energy of the lump.
Substituting eq.~(\ref{eq:k-lump_solution}) into eq.~(\ref{eq:DW_Hamiltonian}),
we obtain the energy of BPS $k$-lump configurations as\footnote{
In the case of domain-wall Skyrmions 
in the magnetic field, 
there is an additional term proportional to $|b_{k-1}|^2$ 
in the $k$-lump energy 
due to the full WZW term 
in 
the domain-wall theory. 
By contrast,
there is not such a term in 
eq.~(\ref{eq:lump-energy-on-wall}). 
While the full WZW 
for the rotation is not available yet, 
 we can understand this 
by recalling the origin of such a term 
in the case of magnetic field. 
It comes from 
the topological term $\partial_3 \pi_0$ 
in the WZW term, 
counting chiral solitons. 
It flips its sign, $-\partial_3 \pi_0$,  when 
the orientation of the ${\mathbb C}P^1$ moduli is at the south pole $n_3=-1$, 
while the vacuum of the domain-wall worldvolume theory is 
at the north pole $n_3=+1$. 
Therefore, the center ($n_3=-1$) of a lump costs energy. 
Contrary to this, in the case of rotation, 
the topological term supporting the $\eta$-solitons is 
$\partial_3 \eta$ 
as in eq.~(\ref{eq:CVE}). 
Since all solitons with different ${\mathbb C}P^1$ moduli
have the same boundary condition 
from $\eta=0$ to $\eta=\pi$, 
the topological term 
$\int dx^3 \partial_3 \eta = \pi$  does not depend on the ${\mathbb C}P^1$ moduli, 
in contrast to the case of the magnetic field.
\label{footnote:additional-term}
}
\begin{align}
    E_{\textrm{DW}} 
    &= \left(\frac{4\pi f_{\pi}^2f_{\eta}}{\sqrt{C}} 
    - \mu_{\textrm{B}} 
    \right)k . \label{eq:lump-energy-on-wall}
\end{align}
Thus, when the chemical potential is larger than
 the critical value,
\begin{eqnarray}
    \mu_{\rm B} \geq \mu_c 
    \equiv \frac{4\pi f_{\pi}^2f_\eta}{\sqrt{C}}
    = \frac{4\sqrt{2}\pi f_{\pi}f_\eta}{m_{\pi}}
   \sim 1.21 {\rm [GeV]}, \label{eq:critical}
\end{eqnarray}
the energy of lumps are negative, $E_{\rm DW} \leq 0$ 
so that lumps are spontaneously created. 
This is the domain-wall Skyrmion phase 
for rotation.
In the evaluation in eq.~(\ref{eq:critical}), 
we have used the vacuum values of the physical quantities $f_{\pi}\approx 93\, \textrm{MeV}$ and $m_{\pi}\approx 140\, \textrm{MeV}$, 
and the relation 
$f_{\eta'}/f_\pi = 1.1$ valid for the three flavors,  
assuming that the same 
$f_{\eta}/f_\pi = 1.1$ 
holds for the two flavors. 
The value of $\mu_c$ implies the effective nucleon mass in this environment inside the soliton at finite density under rapid rotation.
It is interesting to note that this value is close enough to the nucleon mass $\sim 938$ MeV in the QCD vacuum.

\section{Summary and discussion}\label{sec:summary}

We have shown the presence of a new phase 
of rapidly rotating QCD matter in high density region, that is  
a domain-wall Skyrmion phase. 
It was previously known based on the chiral Lagrangian with 
the CVE term \cite{Huang:2017pqe} 
that 
 the ground state is a CSL 
consisting of a stack of 
$\eta$-solitons for two flavors
($\eta'$-solitons for three flavors) 
in a high density region under rapid rotation 
\cite{Nishimura:2020odq}.
In a large parameter region, 
a single $\eta$-soliton decays into a pair of 
non-Abelian sine-Gordon solitons \cite{Nitta:2014rxa,Eto:2015uqa}, 
each of which carries 
${\rm SU}(2)_{\rm V}/{\rm U}(1) \simeq {\mathbb C}P^1 
\simeq S^2$ moduli 
as a consequence of the spontaneous breaking 
of the 
vector symmetry ${\rm SU}(2)_{\rm V}$ 
in the vicinity of each soliton 
\cite{Eto:2021gyy}.
In such a non-Abelian CSL, 
we have shown that 
the effective world-volume theory of a single non-Abelian soliton is 
a $d=2+1$ dimensional ${\mathbb C}P^1$ model [O(3) model] with a topological term 
originated from the WZW term, 
eq.~(\ref{eq:EFT}).
We have shown 
that when the chemical potential is larger than 
a critical value in eq.~(\ref{eq:critical}), 
 a lump has negative energy to be spontaneously created on the soliton world-volume, implying the domain-wall Skyrmion phase.
 This lump on the soliton world-volume corresponds to a Skyrmion 
 carrying a baryon number 
 in the bulk point of view, 
 in contrast to the domain-wall Skyrmions in the magnetic field \cite{Eto:2023lyo,Eto:2023wul}, in which case
 one lump on the soliton worldvolume corresponds to two Skyrmions in the bulk. 
 The effective nucleon mass has been found to be 
 $\sim 1.21$ GeV, 
 close to the  nucleon mass $\sim 938$ MeV in the QCD vacuum. 
We have worked out at the leading order ${\cal O}(p^2)$ 
of the ChPT and do not need higher derivative terms 
such as the Skyrme term.

We have shown the presence of 
the domain-wall Skyrmion phase 
at the leading order ${\cal O}(p^2)$ of the 
ChPT.
At this order, the topological lumps are BPS, 
and thus there are no forces between them.
This would not be the case 
if we go to the next leading order 
${\cal O}(p^4)$. 
At this order, 
one has to include four derivative terms 
such as the Skyrme term in ChPT.
The more important is the inclusion of 
the terms proportional to 
the angular velocity 
$\Omega$ 
in the domain-wall effective action 
in eq.~(\ref{eq:EFT0}) that will 
exert centrifugal force on lumps.
It is an important problem 
to investigate
the next leading order  ${\cal O}(p^4)$.

Let us first discuss a possibility that 
 the non-Abelian CSL and domain-wall Skyrmion phase may be reached in the near future low-energy non-central heavy-ion collision experiments.
The magnitude of the largest vorticity of the current experiment is the order of $10^{22}/\textrm{s}$
\cite{STAR:2017ckg,STAR:2018gyt}. 
Although the critical rotation velocity of the $\eta'$-CSL for three flavors is 
larger by one order of magnitude \cite{Nishimura:2020odq},
that of the non-Abelian CSL is smaller than that 
of the $\eta'$-CSL \cite{Eto:2021gyy}. 
Our results can be extended to more realistic case of the three-flavors.
Also, the $\Lambda (\bar{\Lambda})$ hyperon polarization increases as the collision energy $\sqrt{s}$ decreases, implying the larger angular velocity \cite{STAR:2017ckg,STAR:2018gyt}.
Thus, the non-Abelian CSL and and domain-wall Skyrmion phase 
may be reached in the near future low-energy heavy-ion collision experiments.

In this paper, we have not considered the electromagnetism. There are two electromagnetic couplings:
the minimal  and anomalous couplings.
First, 
the non-Abelian CSL is made of the eta meson and neutral pion and 
are neutral in the electromagnetism. 
By contrast, the charged pions have nontrivial profiles around the lumps studied in this paper, 
and thus the lumps are charged. The electromagnetic U(1) symmetry is spontaneously broken in the vicinity of the lumps, and
consequently the lumps are superconducting. 
More precisely, the lumps are superconducting rings, 
so that their sizes are quantized 
if there is an external magnetic field.
Second, there is an anomalous coupling to the electromagnetism 
\cite{Son:2004tq,Son:2007ny}.
Consequently, magnetizations
appear on the non-Abelian CSLs through the topological term
under an external magnetic field $\bm{B}$ 
\cite{Son:2004tq,Son:2007ny,Eto:2013hoa}: 
$
{\cal L}_{\rm top} 
=
\frac{q_{\rm u} \mu_{\rm B}}{4\pi^2}
\nabla \phi_+ \cdot \bm{B}+
\frac{q_{\rm d} \mu_{\rm B}}{4\pi^2}\nabla \phi_-
\cdot \bm{B}$. 
The rotation induced ferro(ferri) magnetism was discussed in our previous paper 
\cite{Eto:2021gyy}.
It is an open question how Skyrmions 
considered in this paper affect on the magnetization.

If we introduce the 
isospin chemical potential $\mu_I$, 
there is also 
an anomalous coupling to the neutral pion $\pi_0$ given by  
$\frac{\mu_{\rm B}\mu_{\rm I}}{2\pi^2 f_{\pi}} 
\Omega \cdot \nabla \pi_0$ 
\cite{Huang:2017pqe}, 
giving an another 
topological term 
in addition to $\Omega \cdot \nabla\eta$ 
that we have considered in this paper. 
In this case, 
 both the neutral pion $\pi_0$ and $\eta$ meson 
try to constitute a lattice with different periodicities, but it is impossible because these solitons interact.
Consequently, the ground state is not periodic anymore and is rather a quasicrystal, 
 as discussed in the case of strong magnetic fields \cite{Qiu:2023guy}. 
 In such a case, we still can discuss 
domain-wall Skyrmions 
forcusing on an $\eta$ or $\pi_0$ soliton 
as a constituent of the quasicrystal 
at least when solitons are well 
separated for very rapid rotation.
There is an additional term on the worldvolume theory on non-Abelian $\eta$ soliton due to 
the above topological term 
(see footnote \ref{footnote:additional-term}).
Then, the lumps would have constraint as the case 
of those for the strong magnetic field 
\cite{Eto:2023lyo,Eto:2023wul}.

While we have considered two flavors in this paper, 
more realistic case is 
three flavors including the strange quark.
In this case, 
the chiral symmetry ${\rm SU}(3)_{\rm L}\times {\rm SU}(3)_{\rm L}$ is spontaneously broken 
to SU$(3)_{\rm V}$
as well as the axial U$(1)_{\rm A}$ symmetry, 
and the order parameter manifold is 
$[{\rm SU}(3)_{\rm L}\times {\rm SU}(3)_{\rm R} 
\times {\rm U}(1)_{\rm A}]/ 
[{\rm SU}(3)_{\rm V} \times {\mathbb Z}_3]\simeq {\rm U}(3)$. 
Then, 
a single U$(3)$ non-Abelian soliton 
spontaneously  breaks SU$(3)_{\rm V}$ to 
its subgroup SU(2) $\times$ U(1), and thus there appear 
${\rm SU}(3)/[{\rm SU}(2) \times {\rm  U}(1)]\simeq {\mathbb C}P^2$ NG modes in the vicinity of the soliton. 
Then, ${\mathbb C}P^2$ lumps 
on the soliton 
corresponds to 
SU$(3)$ Skyrmions in the bulk 
\cite{Eto:2021gyy}.

Now, we make comments on the domain-wall Skyrmions in quark matter at large $\mu_{\textrm{B}}$.
The ground state in the two-flavors case  without rotation is
 the two-flavor superconducting (2SC) phase \cite{Alford:2007xm}
in which the  chiral symmetry is unbroken. 
Instead, for the three flavors, 
the ground state without rotation is 
the color-flavor locked (CFL) phase 
\cite{Alford:1997zt,Alford:1998mk} 
(see ref.~\cite{Alford:2007xm} as a review), in which 
the chiral symmetry ${\rm SU}(3)_{\rm L}\times {\rm SU}(3)_{\rm R}$ is spontaneously broken 
as well as U$(1)_{\rm A}$.
As the case of three-flavor nuclear matter, 
an $\eta'$-CSL phase appears 
under the rapid rotation \cite{Nishimura:2020odq}, 
in which a continuity to the three-flavor CSL in nuclear matter was also studied.
The instanton effect is suppressed due to the Debye screening \cite{Gross:1980br,Shuryak:1982hk}
in the large $\mu_{\textrm{B}}$ region, 
implying small $\beta$ 
so that non-Abelian CSL is favored.
Then, domain-wall Skyrmions can be constructed 
as ${\mathbb C}P^2$ lumps 
on a U$(3)$ non-Abelian soliton. 
It was discussed in ref.~\cite{Hong:1999dk} 
that Skyrmions in the CFL phase 
 can be regarded as quarks instead of baryons 
and are called qualitons.
Thus, quarks condensed outside the non-Abelian solitons which is in the CFL phase 
may not be condensed inside the non-Abelian solitons, 
similar to Andreev bound states in superconductors.

Apart from application to QCD, 
there are also interesting points for 
physics of topological solitons. 
For instance, 
more general 
non-BPS solutions of the ${\mathbb C}P^{N-1}$ model can be constructed by 
the Din and Zakrzewski's projection method  
\cite{Din:1980jg,Din:1980wh,Misumi:2016fno}. 
One of questions is what these correspond to 
in the bulk. 
The other is fractional ${\mathbb C}P^{N-1}$ 
lumps in a twisted boundary condition \cite{Eto:2004rz,Eto:2006mz}. What is the meaning of 
fractional baryons in the bulk perspective?

\begin{acknowledgments}
This work is supported in part by 
 JSPS KAKENHI [Grants No. JP19K03839 (ME) 
No. JP21H01084 (KN), 
 and No. JP22H01221 (ME and MN)] and the WPI program ``Sustainability with Knotted Chiral Meta Matter (SKCM$^2$)'' at Hiroshima University (KN and MN).
\end{acknowledgments}

\bibliographystyle{jhep}
\bibliography{reference.bib}


\end{document}